\documentclass[a4paper,12pt]{article}

\usepackage{epsfig}

\newfont{\sfb}{cmssbx10 scaled 1400}
\newfont{\bigsf}{cmssbx10 scaled 1600}

\begin{document}
\begin{center}
{\LARGE\bf Diffractive $\Lambda_c^+$ Productions in Polarized $pp$ 
Reactions and Polarized Gluon Distribution}\\

\vspace{3.5em}

N. I. Kochelev
\footnote{kochelev@thsun1.jinr.ru}\\
\vspace{0.8em}
{\it Bogoliubov Laboratory of Theoretical Physics,}\\
{\it Joint Institute for Nuclear Research,}\\
{\it 141980 Dubna, Moscow region, Russia}\\
\vspace{1em}
and\\
\vspace{1em}
T. Morii
\footnote{morii@kobe-u.ac.jp}
 and S. Oyama
\footnote{satoshi@radix.h.kobe-u.ac.jp}\\
\vspace{0.8em}
{\it Faculty of Human Development and}\\
{\it Graduate School of Science and Technology,}\\
{\it Kobe University, Nada, Kobe 657-8501, Japan}\\
\vspace{6.5em}
{\bf Abstract}
\end{center}

\baselineskip=22pt
To test the model of the polarized gluon 
distribution $\Delta G(x, Q^2)$ in the
proton, we propose a new process, diffractive $\Lambda_c^+$ productions
in polarized $pp$ reactions, which will be observed in the forthcoming
RHIC and also the proposed HERA-$\vec {\rm N}$ experiments.
The spin correlation between the target proton and the 
$\Lambda_c^+$ produced in the target fragmentation region largely 
depends on $\Delta G(x, Q^2)$ and thus, the process is quite promising 
for testing the models of $\Delta G(x, Q^2)$.

\vspace{2.0em}

PACS number(s): 13.88.+e, 13.85.Ni, 14.20.Lq

\vfill\eject

\baselineskip 22pt
\noindent
In these years, spin structure of nucleons has been one of 
the most challenging topics in nuclear and particle physics\cite{Lampe}. 
Since the surprising observation of the polarized structure function 
of the proton, $g_1^p(x)$, by the EMC collaboration in 1988, 
much progress has been attained theoretically and experimentally in the 
study of the spin structure of nucleons.
A large amount of data on polarized structure functions of proton,
neutron and deuteron were accumulated by many experimental groups such as
SMC\cite{SMC} at CERN and E142\cite{E142}, E143\cite{E143}, E154\cite{E154}, 
E155\cite{E155} at SLAC and HERMES\cite{HERMES} at DESY.
The progress in the data precision is also remarkable.  On the other hand,
the next-to-leading order QCD calculations of polarized splitting 
functions\cite{MvNV} stimulated the theoretical 
activities for studying the polarized
parton distributions in the nucleon.   
Several parameterization models of 
polarized parton distribution functions
which fit well to the data have been 
proposed\cite{GS96, GRSV96, ABFR, BBPSS} and analyses have been 
developed with increasing new data\cite{LSS, GGR}.  
These experimental and theoretical
developments brought about a deep understanding on the behavior of
the polarized parton distribution in the proton.
However, a knowledge of the polarized
gluon distribution which is expected to play an important role 
in the so-called
nucleon spin puzzle is still poor, though many processes have been 
proposed so far to extract information of it.  

In this work, in order to extract information on 
the polarized gluon distribution,
$\Delta G(x, Q^2)$, we propose a different process, i.e. diffractive 
$\Lambda_c^+$ productions in polarized $pp$ collisions at high
energies.  Diffractive process which can be described by the 
Pomeron exchange is also
an interesting current topic\cite{Preda}.  There have been many 
discussions on Pomeron interactions in these years, largely 
motivated by recent HERA experiments.
To study this process is now quite timely because
RHIC will start soon and
one of the main purposes of RHIC experiments\cite{RHIC} is 
to extract the polarized
gluon distribution in the nucleon from various reactions sensitive to
$\Delta G(x, Q^2)$ in the proton.  We expect that RHIC 
will also observe our proposed reactions, the diffractive
$\Lambda_c^+$ production.  The same process will 
be observed in the proposed HERA-$\vec{\rm N}$ 
experiments\cite{HERAN}, too.

The process which we are considering here is
\begin{equation}
p(p_1)+\vec p(p_2)\longrightarrow
p(p_1')+\vec \Lambda_c^+(p_{\Lambda_c})+X,
\label{eqn:E1}
\end{equation}
where particles with arrows indicate that they are longitudinally 
polarized and
parameters in parentheses denote the momenta of respective particles.
The lowest order Feynman diagram for this process is shown
in Fig.1.

Let us start by describing why we are interested in $\Lambda_c^+$
productions. It is well-known that the
$\Lambda_c^+$ is composed of a heavy quark $c$ and
antisymmetrically combined light $u$ and $d$ quarks, and thus, the spin of
$\Lambda_c^+$ is basically originated from the $c$ quark.  In addition,
a $c$ quark is produced from protons just through gluon fusion 
in the lowest order as
shown in Fig.1 and hence, gluon polarization affects the spin of
$\Lambda_c^+$ via $c$ quarks produced from gluons.  In other
word, measurement of the spin of $\Lambda_c^+$ gives us an
information on the polarized gluon in the proton\cite{KMY}.

The cross section of this process can be calculated based on the 
parton model by using the Pomeron model describing a 
diffractive mechanism and the model of polarized gluons, $\Delta G(x, Q^2)$,
and polarized fragmentation functions, $\Delta D_{\Lambda_c^+/c}(\xi)$, 
of an outgoing $c$ quark decaying into a polarized $\Lambda_c^+$ with
momentum fraction $\xi$.
One of the conventional models of Pomeron interactions is the one proposed
by Donnachie and Landshoff(DL)\cite{DL}, in which the Pomeron is considered
to behave like a $C=+1$ isoscalar photon.  
Here we take this model as our
first analysis. As a typical model of the polarized gluon distributions,
we take GS96\cite{GS96} and GRSV96\cite{GRSV96} parameterizations
among many models, since both of them reproduce
well the experimental data on the polarized structure function
of nucleons and thus, seem plausible.  
As for the polarized fragmentation function
$\Delta D_{\Lambda_c^+/c}(\xi)$, unfortunately it is not known 
at present because of lack of 
experimental data, though we have some knowledge of the unpolarized
function $D_{\Lambda_c^+/c}(\xi)$\cite{Peterson}.  However, since
a $c$ quark is heavy and hence, it is not expected to change much 
its spin alignment during the fragmentation process, it might not be 
unreasonable to use $D_{\Lambda_c^+/c}(\xi)$ for 
$\Delta D_{\Lambda_c^+/c}(\xi)$.
Here we use the model of Peterson et al.\cite{Peterson}
as a substitute for $\Delta D_{\Lambda_c^+/c}(\xi)$.
 
To calculate the cross section of this process
based on the parton model, it is convenient to use
the scaling variables,
$x=\frac{-\Delta^2}{2 p_2\cdot \Delta}$,
$y=\frac{p_2 \cdot \Delta}{p_2 \cdot p_1}$ and 
$z=\frac{p_2 \cdot p_{\Lambda_c}}{p_2 \cdot \Delta}$,
where $\Delta =p_1-p_1^{\prime}$ is the momentum transfer of the
incoming unpolarized proton.
The subprocess of this scattering is
\begin{equation}
 p(p_1) + g (k) \longrightarrow p(p_1') + c(p_c) + \bar c(p_{\bar c}),
\label{eqn:E2}
\end{equation}
where $g$ and $c(\bar c)$ represent the gluon and
$c$-quark($\bar c$-quark), respectively.  Momenta of individual particles
are given in parentheses and are related to the ones in the physical
process of eq.(\ref{eqn:E1}) as $k=\xi p_2$, 
$p_c=\frac{p_{\Lambda_c}}{\xi'}$,
where $\xi$ and $\xi'$ are momentum fractions of the gluon to the
target proton and the $\Lambda_c^+$ to the $c$-quark, respectively.
For this subprocess, we can also define the scaling variables,
$x_p=\frac{-\Delta^2}{2 k \cdot \Delta} = \frac{x}{\xi}$, 
$y_p=\frac{k \cdot \Delta}{k \cdot p_1} = y$ and  
$z_p=\frac{k \cdot p_c}{k \cdot \Delta} = \frac{z}{\xi'}$.
Here we follow the Schuler's way\cite{Schuler} to calculate
the 3-body phase space
for the subprocess of eq.(\ref{eqn:E2}).  
Then, we define the particle momenta in the final state $c\bar c$
quark(or Pomeron-gluon) C.M.S. in the subsystem (\ref{eqn:E2}),
with $\vec p_c + \vec p_{\bar c} = \vec\Delta + \vec k = 0$,
where the gluon momentum $\vec k$ and the target proton momentum
$\vec p_2$ point to the positive $z$ direction.  An angle $\phi$ 
between the proton plane $(\vec k\times \vec p_1)$
and the $c$-quark plane $(\vec k\times \vec p_c)$   
is defined by
$\cos\phi = \frac{(\vec k\times \vec p_1)\cdot(\vec k\times \vec p_c)}
{|\vec k\times \vec p_1||\vec k\times \vec p_c|}$.
In addition to $s=(p_1+p_2)^2$, it is convenient
to define the Lorentz invariant kinematical variables,
$\hat s=(k + \Delta)^2$, 
$\hat t_1=(p_1 - p_1')^2=\Delta ^2$ and  
$\hat t_2=(k - p_c)^2$,
which for $s\gg m_p^2$, can be expressed in terms of $x$, $y$, and $z$ 
as $\hat s=(\xi -x)ys$, 
$\hat t_1=-xys$ and  
$\hat t_2=-\frac{\xi}{\xi'}yzs+m_c^2$, respectively.

By using these variables, we can calculate the differential
cross section for the process of eq.(\ref{eqn:E1}) as follows;
\begin{equation}
 \frac{d \Delta\sigma}{dy dz}=\int_{x_{\min}}^{x_{\max}}d x
\int_{\xi_{\min}}^1 \frac{d \xi}{\xi}
\int_{\xi'_{\min}}^1 \frac{d \xi'}{\xi'}\int_0^{2\pi} d\phi
\Delta G(\xi, Q^2)
\frac{d \Delta\hat\sigma}{dx dy dz d\phi}\Delta D_{\Lambda_c^+/c}(\xi'),
\label{eqn:E3}
\end{equation}
where the kinematical limits of the integral variables are given as
$x_{\max}=\frac{1}{ys}$,
$x_{\min}=\frac{m_p^2 y}{(1-y)s}$,
$\xi_{\min}=x+\frac{4m_c^2}{ys}$ and 
$\xi'_{\min}=\frac{\hat{s}z}{2m_c^2}
   \left(1-\sqrt{1-\frac{4m_c^2}{\hat{s}}} \right)$; 
$x_{\max}$ is
determined from $-1\leq \hat t_1$ which we take for ensuring
the diffractive condition for the incident 
proton with momentum $p_1$, 
$x_{\min}$ from $\sin\gamma\geq 0$ where $\gamma$ is the angle
between the momentum of the incoming unpolarized proton and
$z$ axis,
$\xi_{\min}$ from the condition 
$\hat s\geq 4m_c^2$ and $\xi'_{\min}$ from the 
requirement $-1\leq \cos\theta \leq 1$ for the scattering angle 
$\theta$ of the final $c$-quark in the subprocess (\ref{eqn:E2}),
respectively.
In eq.(\ref{eqn:E3}), the polarized subprocess cross section
is given as
\begin{eqnarray}
 \frac{d\Delta\hat\sigma}{dx_p dy_p dz_p d\phi}
&=&\frac{d\hat\sigma_{++}}{dx_p dy_p dz_p d\phi}-
\frac{d\hat\sigma_{+-}}{dx_p dy_p dz_p d\phi}+
\frac{d\hat\sigma_{--}}{dx_p dy_p dz_p d\phi}-
\frac{d\hat\sigma_{-+}}{dx_p dy_p dz_p d\phi}, \nonumber \\
&=& \frac{y_p}{512 \pi^4}\frac{1}{2}
(|{\cal M}|^2_{++} - |{\cal M}|^2_{+-}
+ |{\cal M}|^2_{--} - |{\cal M}|^2_{-+}), \nonumber \\
&=&\frac{y_p}{1024 \pi^4}( \Delta|{\cal M}_1|^2+\Delta|{\cal M}_2|^2
     -2{\rm Re}\{\Delta({\cal M}_1{\cal M}_2^*)\}),
\label{eqn:E4}
\end{eqnarray}
with
\begin{eqnarray}
-i {\cal M} &=& -i{\cal M}_1-(-i{\cal M}_2),\\
\Delta|{\cal M}_{1,2}|^2 &=& |{\cal M}_{1,2}|^2_{++} -
|{\cal M}_{1,2}|^2_{+-} +
|{\cal M}_{1,2}|^2_{--} - |{\cal M}_{1,2}|^2_{-+},\\
\Delta({\cal M}_1{\cal M}_2^*) &=& ({\cal M}_1{\cal M}_2^*)_{++} -
({\cal M}_1{\cal M}_2^*)_{+-} +
({\cal M}_1{\cal M}_2^*)_{--} - ({\cal M}_1{\cal M}_2^*)_{-+},
\end{eqnarray}
where $\frac{d\hat\sigma_{+-}}{dx_p dy_p dz_p d\phi}$ and
$|{\cal M}|^2_{+-}$,
for example, denote that the helicity of the gluon and the $c$ quark is
positive and negative, respectively.
${\cal M}_1$ is the amplitude corresponding to the subprocess shown in
the Feynman diagram of Fig.1 and
explicitly written by
\begin{eqnarray}
{\cal M}_1 &=& i\bar u (p_c) \beta \gamma^\mu
        \frac{i(\not k-\not p_{\bar c}+m_c)}{(k-p_{\bar c})^2-m_c^2}
        (-i g_s t^a\not \epsilon)v(p_{\bar c}) \nonumber \\
&&  \times \bar u (p_1') 3 \beta_0 F(t)f((k-p_{\bar c})^2-m_c^2)
\gamma_\mu u(p_1)
        \left( \frac{s_1}{s_0} \right)^{\alpha(t)-1},
\end{eqnarray}
where $\beta$ and $\beta_0$ are the charm quark-Pomeron and 
light quark-Pomeron couplings, respectively. 
$s_1=(p_1'+p_c)^2$ and $s_0$ is a scaling constant fixed as
$s_0=1{\rm GeV}^2$.  ${\cal M}_2$ has similar expression which is 
originated from an 
interchange of $c$ and $\bar c$ in the final state 
in the subprocess.  Note that ${\cal M}_1$ and ${\cal M}_2$ are 
added with negative sign because of positive charge conjugation
of the Pomeron, as shown in eq.(5).  
The unpolarized
cross section can be calculated similarly by replacing the polarized
functions in the integrand of eq.(\ref{eqn:E3}), 
$\Delta G(\xi, Q^2)$, 
$\frac{d \Delta\hat\sigma}{dx_p dy_p dz_p d\phi}$, and 
$\Delta D_{\Lambda_c^+/c}(\xi')$,  
by the unpolarized ones, 
$G(\xi, Q^2)$,
$\frac{d \hat\sigma}{dx_p dy_p dz_p d\phi}$, and 
$D_{\Lambda_c^+/c}(\xi')$, respectively.

The 2-spin asymmetry of $\Lambda_c^+$ is calculated by
\begin{eqnarray}
A_{LL}^{\Lambda_c^+} &\equiv& \frac{d\hat\sigma_{++}-d\hat\sigma_{+-}
                        +d\hat\sigma_{--}-d\hat\sigma_{-+}}
              {d\hat\sigma_{++}+d\hat\sigma_{+-}
                        +d\hat\sigma_{--}+d\hat\sigma_{-+}}
=\frac{d \Delta\sigma}{dy dz} / \frac{d \sigma}{dy dz}.
\end{eqnarray}
By using these formulas, we have calculated 
the polarized and unpolarized cross sections and the 
2-spin asymmetry for the diffractive
$\Lambda_c^+$ production in $p\vec p$ reactions at $\sqrt{s}=50\ {\rm GeV}$
and $\sqrt{s}=500\ {\rm GeV}$,
foreseeing the forthcoming RHIC experiments.

Concerning the parameters related to the Pomeron
interaction, we use the same ones given in Ref.\cite{DL}, such as
$F(t)=\frac{4m_p^2-2.79t}{4m_p^2-t}\frac{1}{(1-t/0.71)^2}$,
$f(p^2-m^2)=\frac{\mu_0^2}{\mu_0^2-(p^2-m^2)}$ with $\mu_0=1.0$GeV.
The trajectory of the Pomeron is given as
\begin{equation}
 \alpha(t) = 1.08 + 0.25 t.
\end{equation}

As for the quark-Pomeron coupling, the light quark-Pomeron coupling
is fixed as 
${\beta_0}^2 \sim 3.5 {\rm GeV}^{-2}$\cite{DL}.
However, it is known from experiment that 
the effective coupling of the Pomeron to
heavy quark is rather weaker than the one of a light quark\cite{DLh}.
Therefore,
we use $\beta \sim (m_{u,d}/m_{c})\beta_0 \sim 0.23 \beta_0$
for the charm quark-soft Pomeron coupling,
by taking account of the mass effect of charm quark propagator
in the Pomeron-charm quark vertex.
Other parameters are fixed as
$m_c = 1.5\ {\rm GeV}$
and $Q^2=(2m_c)^2$.

As mentioned above, for the polarized gluon distribution function
$\Delta G(x, Q^2)$, we take two typical parameterization models of
GS96\cite{GS96} and GRSV96\cite{GRSV96}, while for the 
unpolarized distribution $G(x, Q^2)$, we take the model of 
GRV94\cite{GRV94}.  For the
polarized and also unpolarized fragmentation functions, 
$\Delta D_{\Lambda_c^+/c}(\xi')$ and $D_{\Lambda_c^+/c}(\xi')$, we use
the same function of the model of Peterson et al. taken up by
the Particle Data Group in Ref.\cite{PDG}.

Calculated cross sections and 2-spin asymmetries
are presented in Figs. 2 and 3.
Since the variables $y$ and $s$ are included in the amplitude mostly 
as a factorized form $ys$, the cross sections have 
almost the same behavior for the same values of $ys$.
Since we are interested in the target 
fragmentation regions, we have shown only the case
with $z=0.1$, whose kinematical range mostly covers the region
with positive rapidity for the produced $\Lambda_c^+$.   
Note that we have taken the $z$ axis to be the
direction of the target proton $\vec p_2$ and thus,
the region with positive rapidity for the produced $\Lambda_c^+$
corresponds to the target fragmentation region.  
As shown in Figs.2 and 3, the 2-spin asymmetry
$A_{LL}^{\Lambda_c^+}$ rather largely depends on the model of  
$\Delta G(x, Q^2)$ in some kinematical regions.  
Therefore, the 2-spin asymmetry for the proposed process is
promising for testing the gluon polarization.
Of course, 
the analysis might be rather primitive because the present calculation
is limited in the lowest order.  To get more reliable predictions,
in addition to the next-to-leading order calculation, 
we have to refine the Pomeron model and also to have a good knowledge
of the polarized fragmentation function of a $c$ quark to 
$\Lambda_c^+$ decays.  
Although these subjects have  
their own interest and need further investigation, they 
are out of scope for the present work.

In summary, we have calculated the cross section and 
the 2-spin asymmetry for diffractive
$\Lambda_c^+$ productions in polarized $pp$ reactions for the target
fragmentation region at $\sqrt s=50$GeV and 500GeV.  We found that the
calculated results largely depend on the model of $\Delta G(x, Q^2)$
in some kinematical regions and thus, the process is 
quite promising for extracting information on
polarized gluons in the proton.

Although in this work calculations were carried out expecting the
forthcoming RHIC experiments, the same analysis might be applied also 
for the proposed HERA-$\vec {\rm N}$ experiments.

\newpage
\begin{figure}[t]
\centering
\epsfxsize=8cm
\epsfysize=5cm
\epsfbox{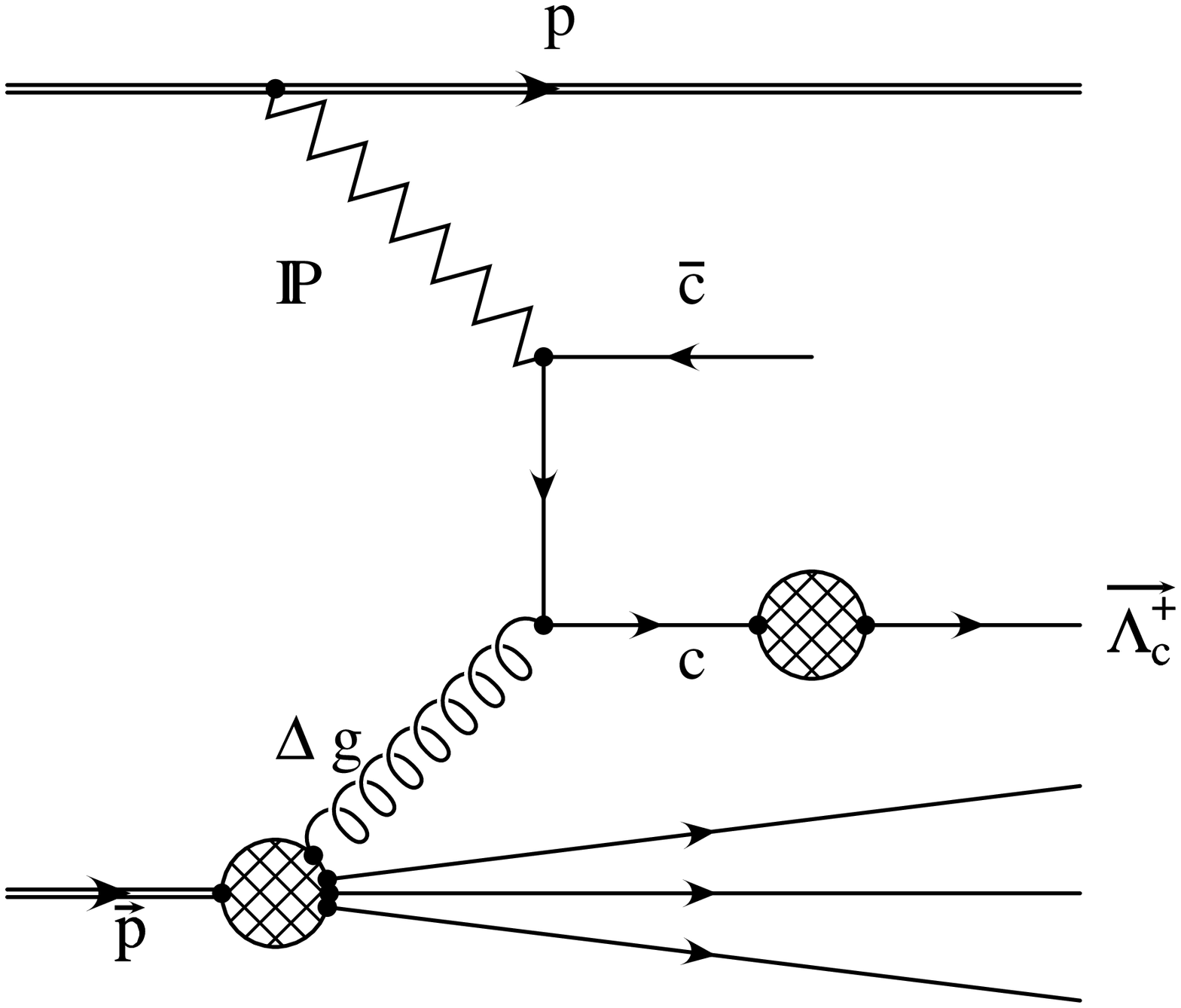}
\caption{Lowest diagram for $p+\vec p\longrightarrow
p+\vec \Lambda_c^++X$.}
\end{figure}
\vspace{-5cm}
\begin{figure}[hb]
\centering
\epsfxsize=7cm
\epsfysize=5cm
\epsfbox{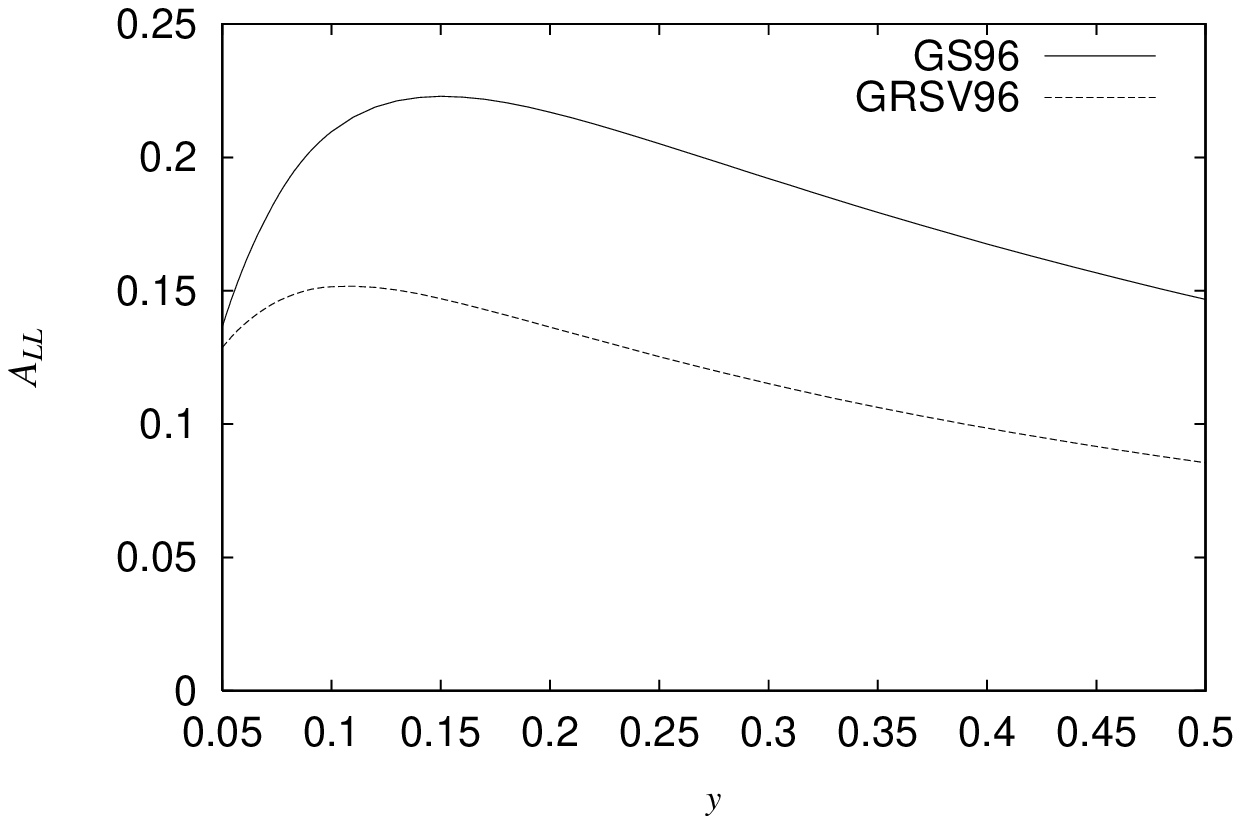}
\hspace{1cm}
\epsfxsize=7cm
\epsfysize=5cm
\epsfbox{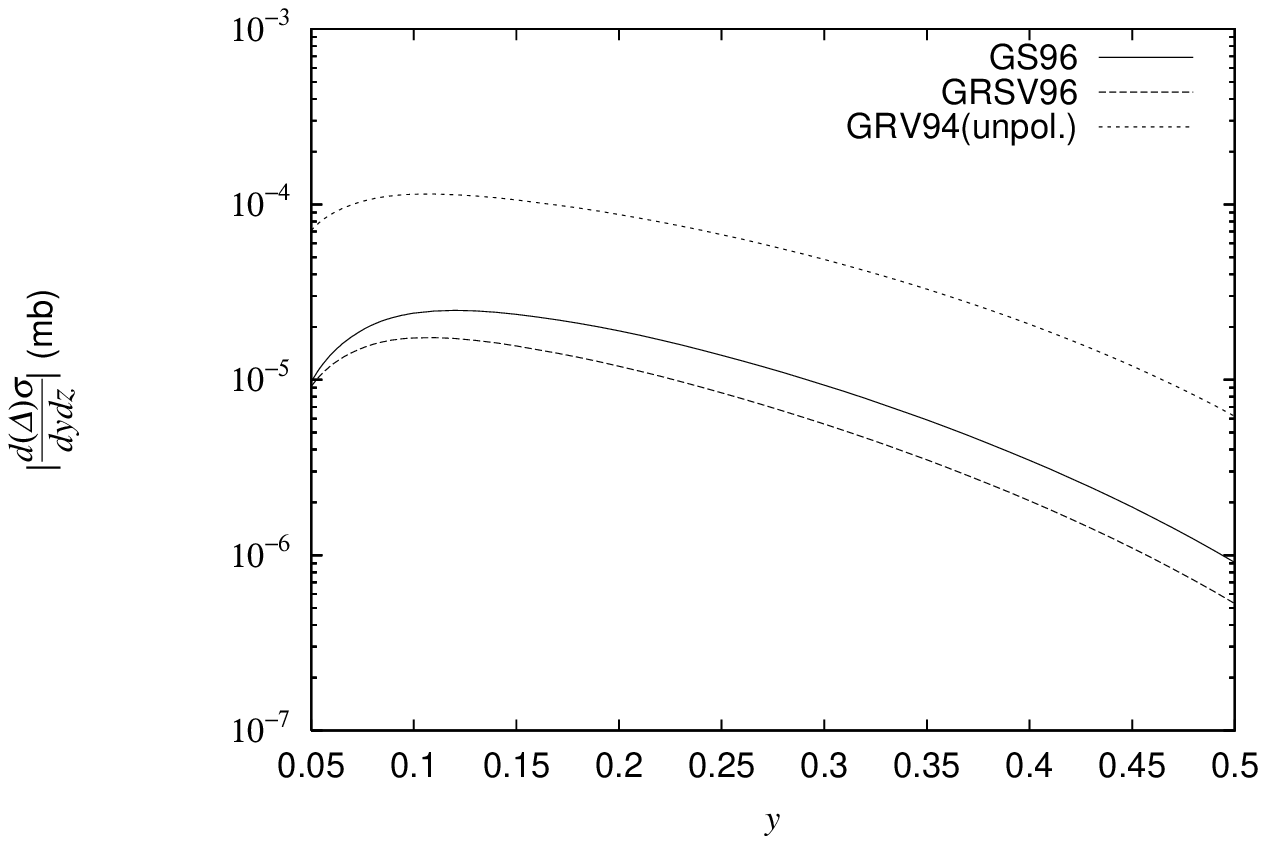}
\vspace{-5mm}
\caption{2-spin asymmetry and cross sections for $\sqrt s=50{\rm GeV}$
and $z=0.1$.
Solid line is for GS96 and dashed for GRSV96. In the right figure,
unpolarized cross section is also given in dotted line.
}
\vspace{1cm}
\centering
\epsfxsize=7cm
\epsfysize=5cm
\epsfbox{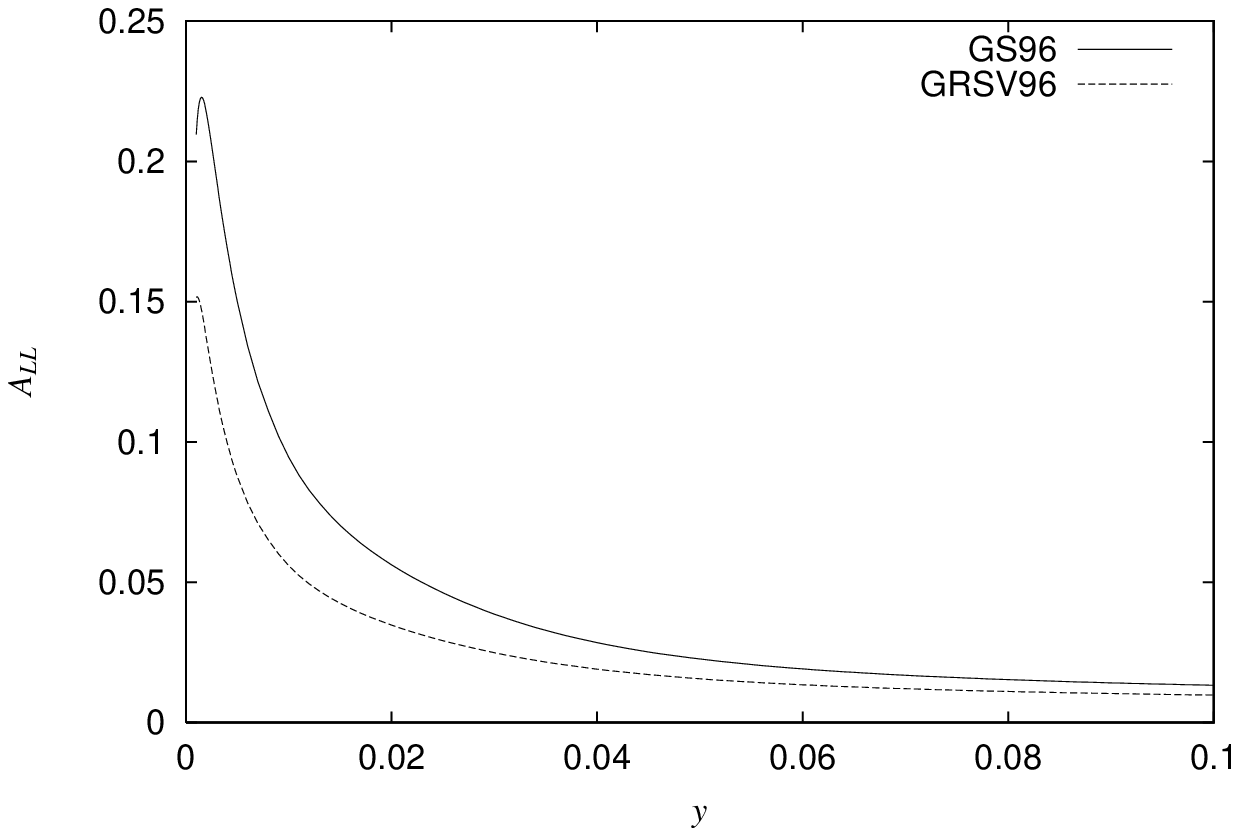}
\hspace{1cm}
\epsfxsize=7cm
\epsfysize=5cm
\epsfbox{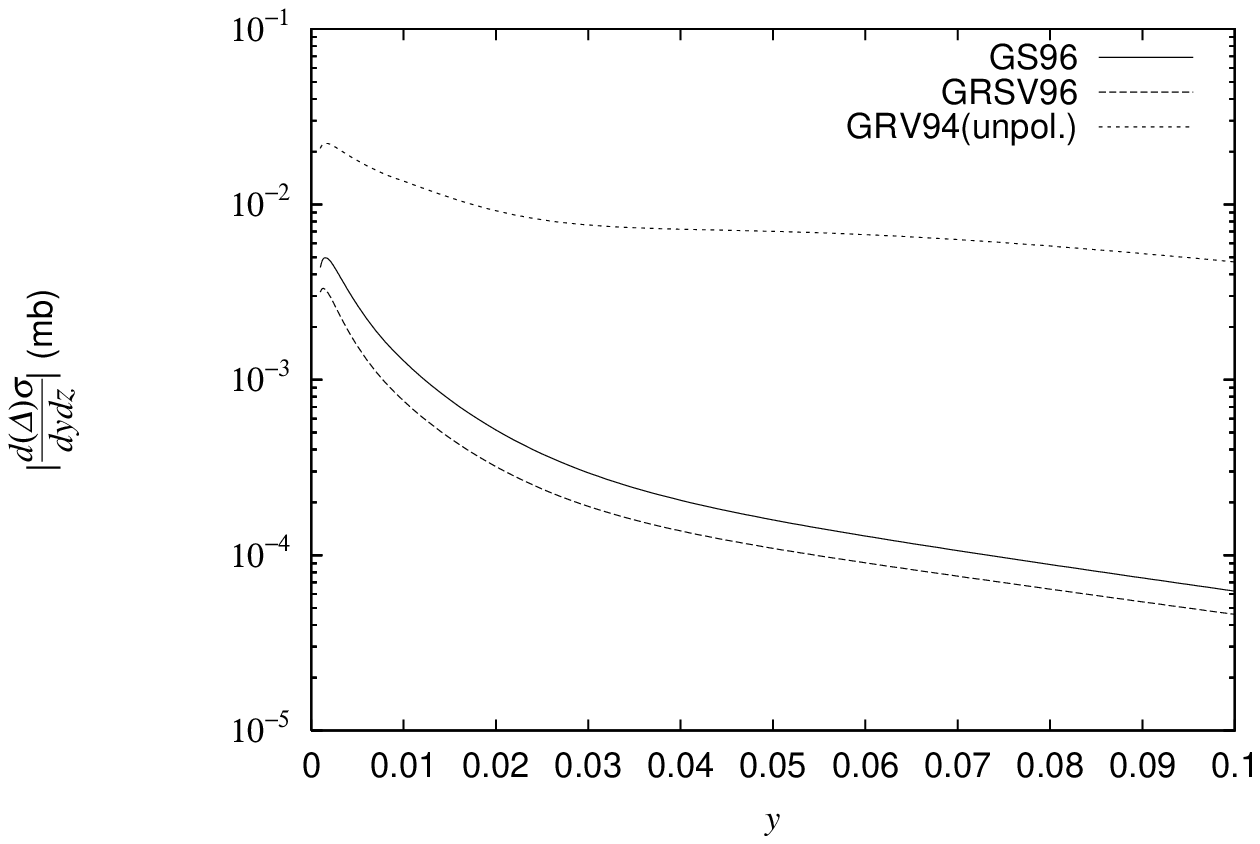}
\vspace{-5mm}
\caption{
The same as in Fig. 2, but for $\sqrt s=500{\rm GeV}$.
}
\end{figure}
\end{document}